\documentclass[12pt]{iopart}
\usepackage{graphicx}
\usepackage{color}
\usepackage{latexsym}

\begin{document}
\title{Effect of differential cross section in Breit-Wheeler pair beaming}

\author{X Ribeyre, E d'Humi\`eres, S Jequier and V T Tikhonchuk}
\address{Univ. Bordeaux-CNRS-CEA, Centre Lasers Intenses et Applications, UMR 5107, 33405 Talence, France}

\ead{ribeyre@u-bordeaux.fr}

\begin{abstract}
The pair beaming in the Breit-Wheeler (BW) process is investigated. We examine the effect of the BW differential cross section on pair angular and energy distributions. Although, this study is relevant for laser induced intense gamma-ray source collisions experiments, we apply the pair beaming in astrophysical context, in particular for Active Galactic Nuclei (AGN).
\end{abstract}

\noindent{ Keywords:\/ gamma-ray sources, Breit-Wheeler process, pair electron-positron, Active Galactic Nuclei}
\\
\\
\maketitle

\section{Introduction}
Electron-positron pair production in the collision of two photons is a fundamental physical process predicted by the quantum electrodynamics (QED)~\cite{QED_Book}. The electron-positron pair annihilation in two photons: $e^++ e^- \rightarrow \gamma+\gamma$ was theoretically predicted by Dirac~\cite{Dirac_1930} and has obtained extremely accurate experimental verification~\cite{Klemperer_1934}.  But the reverse process: $\gamma+\gamma \rightarrow e^+ + e^-$, theoretically predicted by Breit-Wheeler (BW)~\cite{Breit_1934}, although conceptually simple, has been by far more difficult to verify experimentally. 

The photon-photon collision physics was first studied in 1970's in Novosibirsk~\cite{Novosibirsk_1971} and Frascati~\cite{Frascati_1979}. In these experiments the pairs were created in the collision of electron and positron beams: $e^++e^-\rightarrow e^++e^-+e^++e^-$ in the so-called Landau-Lifshitz process. The photons in this process are virtual, they present an intermediate step of the pair creation: $\gamma^*+\gamma^* \rightarrow e^++e^-$. Another similar QED process verified experimentally is the Bethe-Heitler process~\cite{Anderson_1933}, $\gamma+\gamma^*\rightarrow e^++e^-$, where a real photon collides with a virtual photon produced in the Coulomb electric field of an ion. Concerning the collision of real photons only the non-linear or multi-photon BW process~\cite{Reiss_1962} has been observed~\cite{Burke_1997}, where a real photon is coupled to several (4-5) low energy laser photons. The two photons collision process is still awaiting a direct observational verification. Only recently, thanks to 
the high power laser technology, first scheme for its direct observation have been proposed~\cite{Pike_2014,Ribeyre_2016,Drebot_2017}.

In the astrophysical context, the BW process plays an important role in high energy phenomena~\cite{Ruffini_2010}. According to Nikishov~\cite{Nikishov_1962}, the background photons put a stringent limit on the maximum energy of photons coming from hard X-ray sources. The BW process also takes place in Active Galactic Nuclei (AGN)~\cite{Bonometto_1971, Li_1996}, where the gamma rays near the central black hole are converted into electron-positron pairs, which are further accelerated by the radiation pressure of the X-ray flux emission from the disk.

Electron-positron pair production by two gamma photon beams was studied in Ref.~\cite{Ribeyre_2017} by considering only the pairs kinematics in the BW process. The pair angular distribution in the center of mass (CM) frame was assumed to be uniform, i.e. the pairs are emitted isotropically. The effect of pair beaming in the laboratory frame was explained by the directed motion of the CM frame. This effect may facilitate the experimental observation of the BW process. 

In this paper, we account for the effect of the BW differential cross section anisotropy on the pair beaming. This effect changes the angular and energy distribution of pairs in the laboratory frame. In Sect.~\ref{Sect1}, we recall the main features of the pair beaming assuming an anisotropic pairs distribution in the CM frame. Section~\ref{Sect2} is devoted to the analysis of the effect of BW differential cross section anisotropy on pair beaming. The BW differential cross section is derived in Sec.~\ref{Sect3}, and the pair angular and energy distributions are studied in Sect.~\ref{Sect4}. Section~\ref{Sect5}, presents an analysis of the BW process in the astrophysical context, in particularly for AGN. Section~\ref{Sect6} contains our conclusions. 

\section{Kinematics of $e^+,e^-$ pairs and pair beaming}\label{Sect1}

In this section we recall the main results obtained in Ref.~\cite{Ribeyre_2017}, were we demonstrated the pair beaming effect in the collision of two photon beams. We use a unit system where the speed of light $c=1$ and the electron mass $m_e=1$. We consider two colliding photons with momenta $\vec p_{\gamma_1}$, $\vec p_{\gamma_2}$ and energies $E_{\gamma_1}$, $E_{\gamma_2}$, respectively and $\theta_p$ is the angle between them. The scalar product in Sec.~\ref{Sect3}, of the four-momenta of two photons is a Lorentz invariant: $2E_{\gamma_1} E_{\gamma_2} (1-\cos\theta_p)=E_{cm}^2$, where $E_{cm}$ is the total energy in the CM frame. The pair creation threshold is defined by the condition $E_{cm}\ge2$, electrons and positions are created at rest if $E_{cm}=2$. In the general case this condition writes: $E_{\gamma_1} E_{\gamma_2} (1-\cos\theta_p)\ge 2$. For a given energy of the two photons, the threshold angle $\theta_{th}$ is defined by: $\theta_p \ge \theta_{th}$ where $\theta_{th}=\arccos(1-1/E_{\gamma_1}
E_{\gamma_2})$. From the Lorentz transformation of the $e^-, e^+$ pair momenta and energy, one finds a condition in which all pair momenta are aligned in the direction of the CM frame velocity. This condition is referred to as the: pair beaming condition. This condition writes: $\theta_p\le\theta_{p,beam}$, where: $$\theta_{p,beam}=\arccos(1-[E_{\gamma_1}+E_{\gamma_2}]/E_{\gamma_1} E_{\gamma_2}).$$ Moreover, in the case where $(E_{\gamma_1}+E_{\gamma_2})/2E_{\gamma_1}E_{\gamma_2}>1$, there is beaming for $\theta_p\ge\theta_{th}$. This latter beaming condition is relevant for collisions between very high and very low energy photons. For $\theta_p=\pi$ and equal photon beam energies, there is no beaming because the laboratory frame and the CM frame are identical, and the pairs $e^-,e^+$ are emitted in opposite directions. When the pair beaming conditions apply, the emission cone aperture $\theta_{e,beam}$ writes: 
\begin{eqnarray}
\tan \theta_{e,beam} = \pm \frac{E_{cm}\sqrt{E_{cm}^2-4}}{\sqrt{4(E_{\gamma_1}+E_{\gamma_2})^2-E_{cm}^4}}. \nonumber
\end{eqnarray}

\begin{figure}
\centerline{\includegraphics[width=12cm]{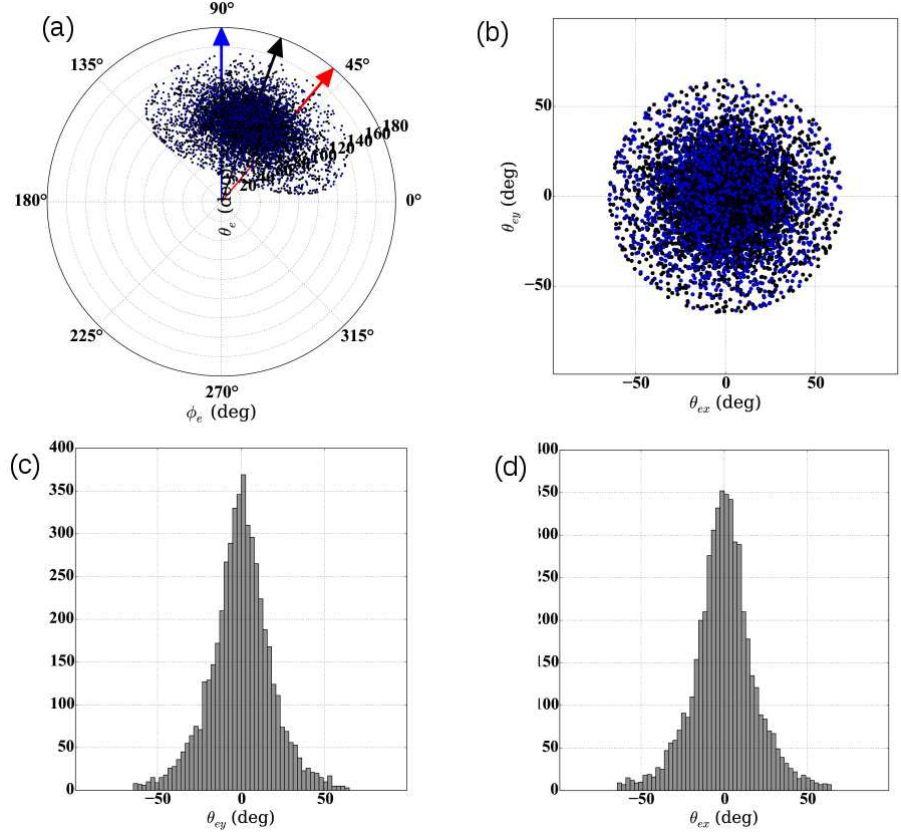}}
\caption{\label{Fig1} Pair emission in the laboratory frame for $E_{\gamma_1}=E_{\gamma_2}=4$ MeV and $\theta_p=40^{\circ}$. (a) Angular distribution of pair emission, $\theta_e$ and $\phi_e$ are the polar and azimuthal angles. Red and blue arrows delimit the pair emission directions, black arrow the direction of CM frame velocity. (b) Pair angular distribution in the direction of the CM frame velocity, blue and black points show the positrons and electrons, respectively. (c) Angular histogram ($\theta_{ex}$) in the plane of incidence of photons. (d) Angular histogram ($\theta_{ey}$) in the direction perpendicular to the incidence plane.}
\end{figure}

However, the results in Ref.~\cite{Ribeyre_2017} are obtained assuming that in the CM frame, the pair emission probability is isotropic, that is, the differential cross section $d\sigma/d\Omega_{cm}=1$ is constant. The pair beaming is then possible for $\theta_{th} < \theta_p<\theta_{p,beam}$. An example of the beaming effect, is shown in Figure~\ref{Fig1} for the case of two photon beams with the same energy: 4~MeV crossing at an angle $\theta_p=40^{\circ}$ which satisfy the beaming condition: $\theta_{p,beam}\simeq 42^{\circ}$, $\theta_{th}\simeq 15^{\circ}$. Five thousands of pairs were emitted isotropically in the CM frame and their distribution in the laboratory frame is shown in Figure~\ref{Fig1}. Figure~\ref{Fig1}a shows the angular pair distribution versus the polar and azimuthal angles ($\theta_e, \phi_e$). The pairs are emitted along the CM frame velocity. The pairs are distributed inside a cone angle $\theta_{e,beam}=\pm 64.7^{\circ}$ (see Fig.\ref{Fig1}b). Figures~\ref{Fig1}c,d present the 
angular distribution in the photon incidence plane and perpendicular to this plane, respectively. The pair distribution has an azimuthally symmetric bell shape. 
\begin{figure}
\centerline{\includegraphics[width=12cm]{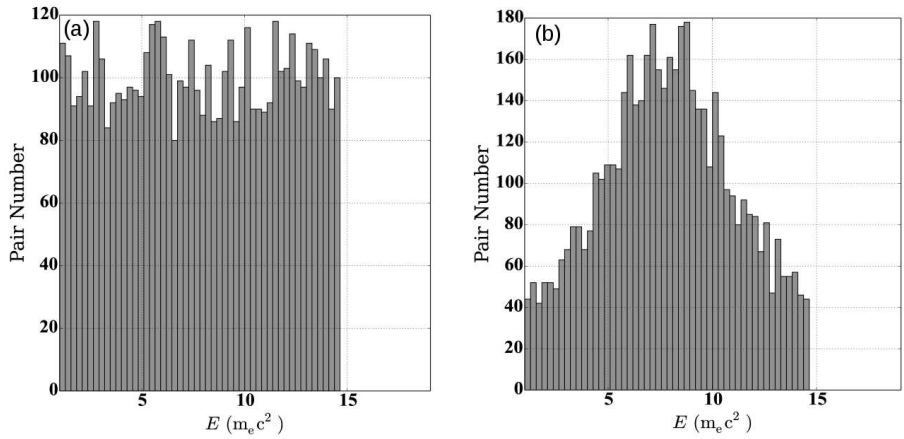}}
\caption{\label{Fig2} Pair energy distribution for $E_{\gamma_1}=E_{\gamma_2}=4$ MeV and $\theta_p=40^{\circ}$. (a) For isotropic (b) for BW differential cross section.}
\end{figure}

According to the Lorentz transformation the minimum and maximum pair energies are \cite{Ribeyre_2017}: 
\begin{eqnarray} E_{max,min}&=&\frac{(E_{\gamma_1}+ E_{\gamma_2})}{2} \nonumber \pm \frac{\sqrt{(E_{\gamma_1}+E_{\gamma_2})^2-E_{cm}^2} \sqrt{E_{cm}^2-4}}{2E_{cm}}.\nonumber 
\end{eqnarray}
The pair energy distribution is presented in Figure~\ref{Fig2}a. The spectrum is uniform between the two energy values given before. The effect of anisotropy of pair emission is discussed in the next section.
\section{Effect of the cross section anisotropy on pair beaming}\label{Sect2}

\subsection{BW differential cross section}\label{Sect3}

The Feynman diagram technique for the differential cross section calculation is described in Ref.~\cite{QED_Book}. Figure~\ref{Fig3} shows the two first order Feynman diagrams of the BW process: the production of an $e^-,e^+$ pair from the collision of two real photons. The total scattering matrix~$\cal M$ is a sum of two scattering matrices $\cal M_{\rm 1}$ and $\cal M_{\rm 2}$ corresponding to the processes shown in Figures~\ref{Fig3}. The module of $\cal M$ averaged over all spin and polarization configurations writes:
\begin{figure}[h]
\centerline{\includegraphics[width=12cm]{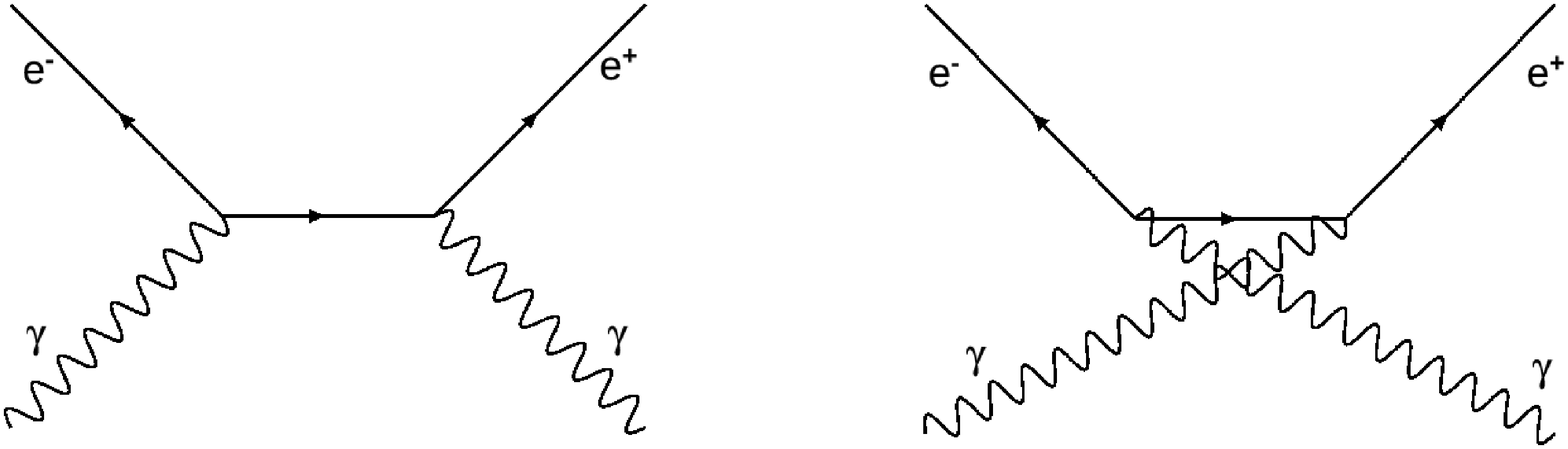}}
\caption{\label{Fig3} Electron-positron pair production from a pair of photons represented to the first order by Feynman diagrams, corresponding to the BW process: $\gamma+\gamma \rightarrow e^+ + e^-$. }
\end{figure}
\begin{eqnarray}
|{\cal M}|^2=\frac{1}{4}\sum_{state}[{\cal  M}_{\rm1}{\cal  M}_{\rm1}^*
+{\cal M}_{\rm2}{\cal M}_{\rm2}^*+2Re({\cal M}_{\rm2}{\cal M}_{\rm1})]. \nonumber
\end{eqnarray}
The pairs are characterized by there four-momenta $P$ and $P^\prime$ and photons by their four-momenta $K$ and $K^\prime$. The kinematic invariants called the Mandelstam variables are then given by: $s=(P+P^\prime)^2=(K+K^\prime)^2$, $t=(P-K)^2=(P^\prime-K^\prime)^2$ and $u=(P-K^\prime)^2=(P^\prime-K)^2$. In particular $\sqrt s=E_{cm}$, is the total energy in the CM frame.

\begin{eqnarray}
|{\cal M}|^2  &=&  2 e^4\left[\frac{u-1}{t^2-1}+\frac{t-1}{u-1} -4\left(\frac{1}{t-1}+\frac{1}{u-1}\right) \right. \nonumber \\
&&\left. -4\left(\frac{1}{t-1}+\frac{1}{u-1}\right)^2\right], \nonumber
\end{eqnarray}
where $e$ the electron electric charge.
The BW differential cross section in the CM frame writes~\cite{Bottcher_1997}:
\begin{eqnarray}
\frac{d\sigma}{d\Omega_{cm}}=\frac{1}{64\pi^2 s}\frac{2|\vec p|}{\sqrt{s}}|{\cal M}|^2, \nonumber
\end{eqnarray}
where, $\vec p$ is the momentum of the positron or electron. 
Expressing the Mandelstam variables through the CM frame velocity $\beta_{cm}$ and the angle between the photon direction and electron emission direction $\theta_x$, one finds:
\begin{eqnarray}
\frac{d\sigma}{d\Omega_{cm}}&=&\frac{r_e^2\beta_{cm}}{s}\left[-1+\frac{3-\beta_{cm}^4}{2} \right.\nonumber\\  
&&\left.\left(\frac{1}{1-\beta_{cm}x}+\frac{1}{1+\beta_{cm}x}\right ) \right. \nonumber \\ 
&&\left. -\frac{2}{s^2}\left(\frac{1}{[1-\beta_{cm}x]^2}+\frac{1}{[1+\beta_{cm}x]^2}\right )\right], \label{Equ1}
\end{eqnarray} where, $r_e$ is the electron classical radius, $x=\cos \theta_x$ and $d\Omega_{cm}=dx d\phi$.
Figure \ref{Fig4}a shows the BW differential cross section in CM frame for the collision of photons with equal energies of 4~MeV and $\theta_p=40^{\circ}$. The energy dependence of the differential cross section is shown in panel (b). 

The probability of pair emission is symmetric along the photons direction axis and achieves a maximum in the photon beam direction, see Fig.~\ref{Fig4}a.  The anisotropy increases with the CM energy. It appears clearly for $\sqrt s>3$, very close to the the pair creation threshold $\sqrt s = 2$. 
\begin{figure}
\includegraphics[width=12cm]{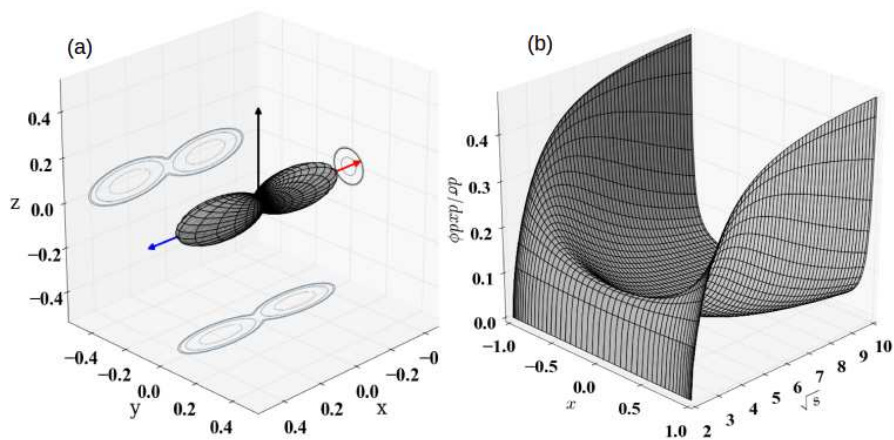}
\caption{\label{Fig4} (a) BW differential cross section $d\sigma/dxd\phi$ (in units of $r_e^2$) blue and red arrows give the photon direction and black arrow the CM frame velocity direction for photon beams with equal energy of 4~MeV and $\theta_p=40^{\circ}$. (b) Dependence of the differential cross section on the photon energy and polar angle.}
\end{figure}
\subsection{Pairs kinematics and energy distributions}\label{Sect4}

\begin{figure}
\centerline{\includegraphics[width=12cm]{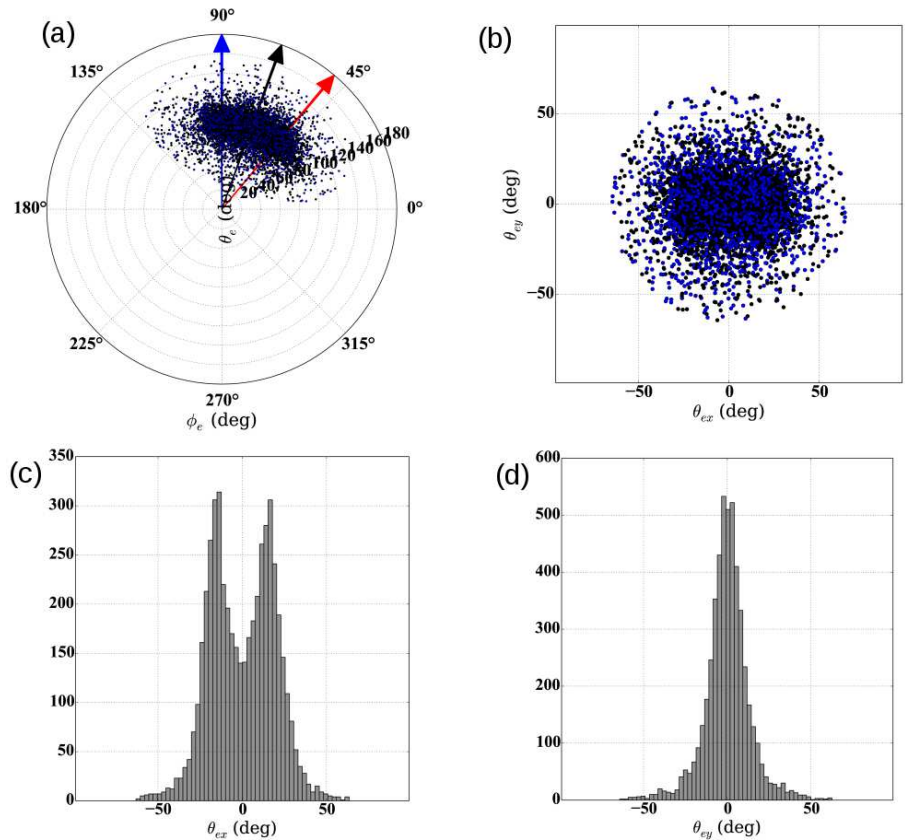}}
\caption{\label{Fig5} Pair emission in the laboratory frame for $E_{\gamma_1}=E_{\gamma_2}=4$ MeV and $\theta_p=40^{\circ}$ with anisotropic differential cross section. (a) Angular distribution of pair emission, $\theta_e$ and $\phi_e$ are the polar and azimuthal angles. Red and blue arrows delimit the pair emission directions, black arrow the direction of CM frame velocity. (b) Pair angular distribution in the direction of the CM frame velocity, blue and black points show the positrons and electrons, respectively. (c) Angular histogram ($\theta_{ex}$) in the plane of incidence of photons. (d) Angular histogram ($\theta_{ey}$) in the direction perpendicular to the incidence plane.}
\end{figure}

By using the Lorentz transformations the pair emission distribution is obtained in the laboratory frame. The pair distribution has been calculated by a random generation of a five thousand pairs in the CM frame with the probability given by Eq.~\eref{Equ1}. Figure~\ref{Fig5} shows the pair emission characteristics for the same set of parameters as in Fig.~\ref{Fig1}, with $E_{cm}=\sqrt s =5.3$. The pair angular distribution is much more affected by the anisotropy of the differential cross section. This is expected as according to Fig.~\ref{Fig4}, the pairs emission is much more anisotropic for $\sqrt s>3$. This indeed can be seen in Figs.~\ref{Fig5}c,d where the pair beam in the collision plane is larger than in the perpendicular direction. In particular, the angular pair distribution shows two peaks in the photon beam direction. The pair energy distribution is shown in in Fig~\ref{Fig2}b. Compared to Fig.~\ref{Fig2}a, the maximum of the energy distribution is shifted to higher energies.

To investigate further the effect of the BW differential cross section on pair beaming, we varied the incoming photon energies. Two different cases are considered: 4~MeV-1~MeV, 4~MeV-7~MeV, while keeping the same crossing angle $\theta_p=40^{\circ}$. For the two cases the values of $\sqrt s$ are: $\sim$3 and $\sim $7 corresponding to an decreasing and increasing anisotropy respectively according to the Fig.~\ref{Fig4}b. 

\begin{figure}
\centerline{\includegraphics[width=12cm]{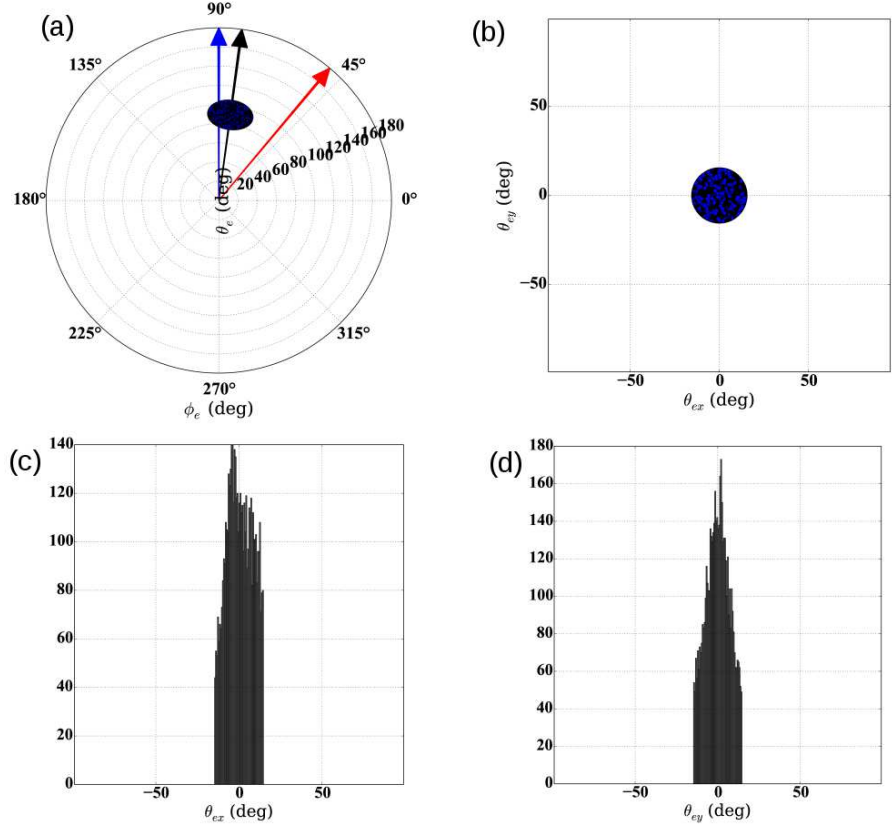}}
\caption{\label{Fig6} Pair emission in the laboratory frame for $E_{\gamma_1}=4$~MeV, $=E_{\gamma_2}=1$~MeV and $\theta_p=40^{\circ}$ with anisotropic differential cross section. (a) Angular distribution of pair emission, $\theta_e$ and $\phi_e$ are the polar and azimuthal angles. Red and blue arrows delimit the pair emission directions, black arrow the direction of CM frame velocity. (b) Pair angular distribution in the direction of the CM frame velocity, blue and black points show the positrons and electrons, respectively. (c) Angular histogram ($\theta_{ex}$) in the plane of incidence of photons. (d) Angular histogram ($\theta_{ey}$) in the direction perpendicular to the incidence plane.}
\end{figure}

\begin{figure}
\centerline{\includegraphics[width=12cm]{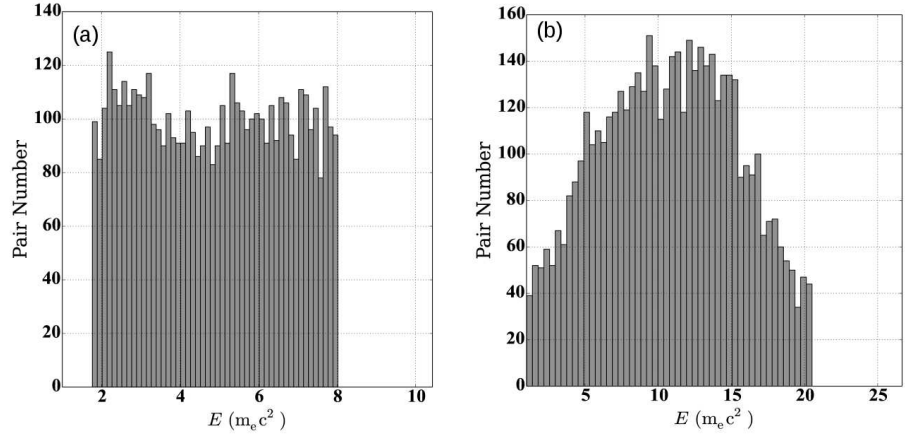}}
\caption{\label{Fig7} Pair energy distribution for $\theta_p=40^{\circ}$. (a) For $E_{\gamma_1}=4$~MeV,$E_{\gamma_2}=1$ MeV and (b) for $E_{\gamma_1}=4$~MeV, $E_{\gamma_2}=7$~MeV.}
\end{figure}

Figure~\ref{Fig6} shows the pair emission characteristics for 4~MeV and 1~MeV photon beams collision.
The pair angular distribution (see Fig.~\ref{Fig6}a) shows a stronger beaming than in Fig.~\ref{Fig5}, indeed, the pair beaming condition gives: $\theta_{e,beam}=\pm 14.7^{\circ}$. Moreover, because of the unbalanced photon energy beams, the CM frame velocity is almost aligned along the high energy photon beam direction. Then most of the pairs are emitted in the direction of the highest photon beam energy. The pairs energy distribution is shown in in Fig~\ref{Fig7}a. Compared to Fig.~\ref{Fig2}b, the maximum of the energy distribution is shifted to higher energies and uniform as in Fig.~\ref{Fig2}a.

\begin{figure}
\centerline{\includegraphics[width=12cm]{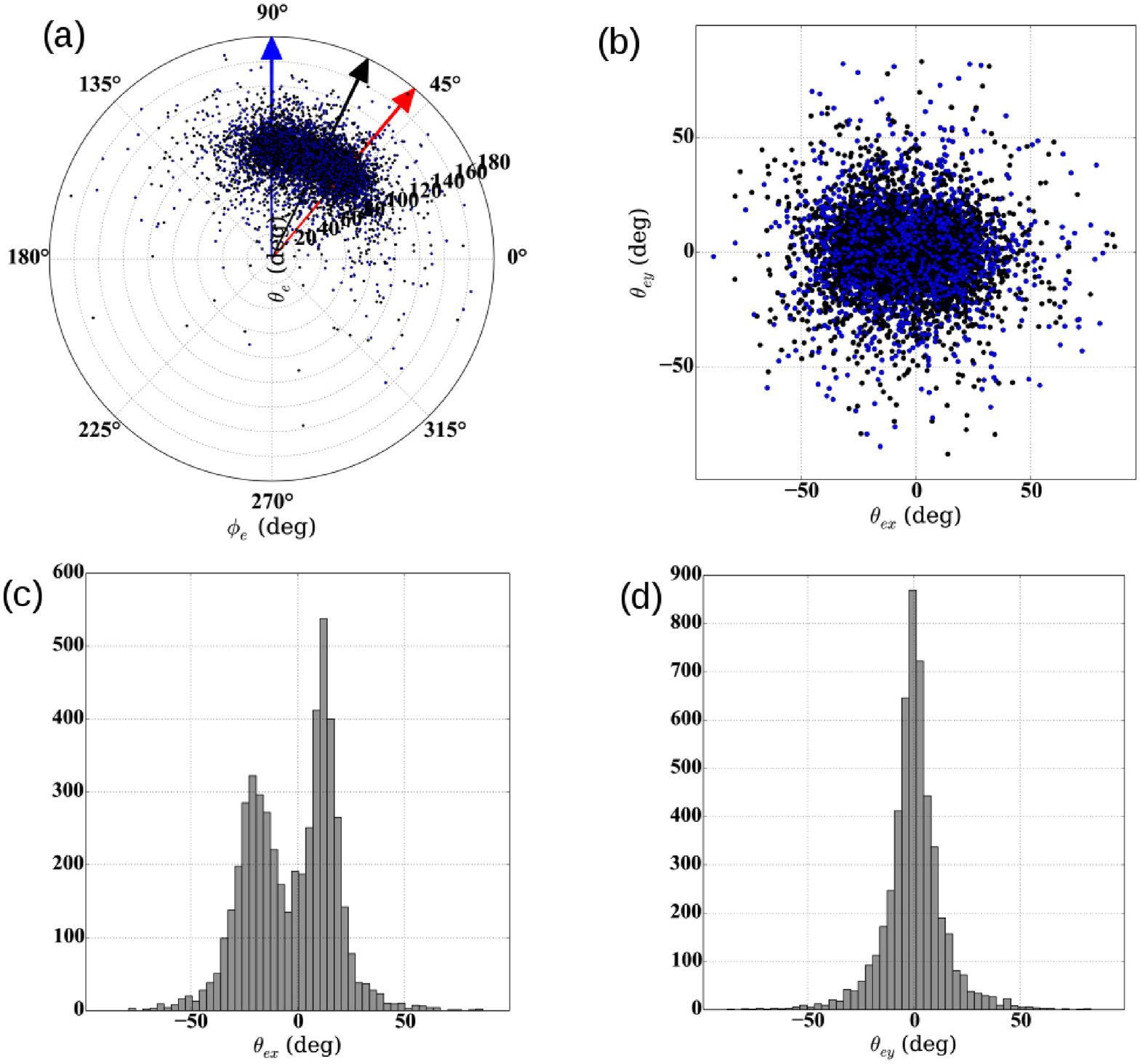}}
\caption{\label{Fig8}  Pair emission in the laboratory frame for $E_{\gamma_1}=4$~MeV, $=E_{\gamma_2}=7$~MeV and $\theta_p=40^{\circ}$ with anisotropic differential cross section. (a) Angular distribution of pair emission, $\theta_e$ and $\phi_e$ are the polar and azimuthal angles. Red and blue arrows delimit the pair emission directions, black arrow the direction of CM frame velocity direction. (b) Pair angular distribution in the direction of the CM frame velocity, blue and black points show the positrons and electrons, respectively. (c) Angular histogram ($\theta_{ex}$) in the plane of incidence of photons. (d) Angular histogram ($\theta_{ey}$) in the direction perpendicular to the incidence plane.}
\end{figure}

For 4~MeV and 7~MeV photon energies, the pair emission characteristics are show in Fig.~\ref{Fig8}. The angular distribution presented in Fig.~\ref{Fig8}a, shows that the CM frame velocity direction is aligned toward the high energy photon direction. Because, the beam angle $\theta_{p,beam}=37^{\circ}$, the beaming condition is not totally satisfied, some pairs are emitted backward compared to the CM frame velocity direction. Figs.~\ref{Fig8}b,c,b show that the pair beam is split in two parts as in Fig.\ref{Fig5}c. However, the pair beam presents an asymmetry in the photon incidence plane and more pairs are emitted in the direction of the highest photon energy beam. The pair energy distribution is shown in Fig~\ref{Fig7}b. The shape of the energy distribution shown in Fig.~\ref{Fig2}b is close to the energy distribution plotted in Fig.~\ref{Fig2}b, but shifted to higher energies.

In summary, to investigate the BW differential cross section effect on pair beaming, three cases are considered in terms of photon beam energies: 4-4~MeV, 4-1~MeV, 4-7~MeV, all cases with $\theta_p=40^{\circ}$. In the case 4-4~MeV, the influences of the differential cross section are important: on pair angular distribution as well as on the energy distribution. The pairs are emitted mainly in the photon beam directions. Then, less pairs are emitted in the bisector between the photon beam directions compared to the case of the isotropic differential cross section. The pair energy distribution achieved a maximum at the mean energy of the two photon beams. Concerning the two last cases: 4-1 MeV; 4-7 MeV, the pair beam is emitted mainly along the CM frame velocity direction. Because of the photon beams energy difference, less pairs are emitted on the bisector between the two photon beams. However, for of 4-1 MeV photon beams collision, the effect of BW differential cross section on the angular pair distribution 
and energy distributions are weak. In the case 4-7 MeV photon beams, the effect becomes important and the pair distribution is beamed in the direction of the most energetic photon beam. The energy distribution is shifted to higher energies and achieves a maximum for the mean energy of the two photon beams.  

\section{Application on pair production in Active Galactic nuclei}\label{Sect5}

\begin{figure}
\centerline{\includegraphics[width=12cm]{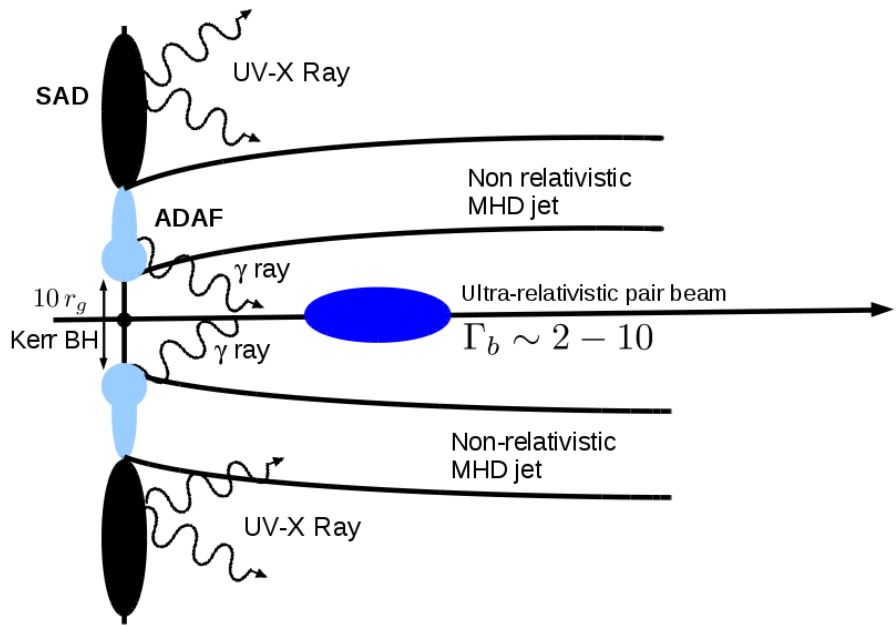}}
\caption{\label{Fig9} Schematic picture of the rotating BH accretion disk axis along with the emission from the inner ADAF and outer SAD. The ADAF emits in the gamma-ray range and the SAD in the UV and X-ray range ($r_g$ is the Schwarzschild radius). The two source flows are represented: a non relativistic outflow from the ADAF driven by the opened magnetic field, and plasma pairs that escape with relativistic speed along the inner flux tubes.}
\end{figure}

In this section we present an analysis of the BW process in the astrophysical context, in particular in AGN. An AGN produces jets of relativistic particles~\cite{Vuillaume_2015}. The relativistic flows are characterized by their bulk Lorentz factor $\Gamma_b$ (equivalent to $E$ on the pair energy distribution see the previous section). The observations show that in AGN the Lorentz factor for jet flows is in range: $\Gamma_b=2-10$. It is mainly assumed that the relativistic flow is dominated by a $e^-,e^+$ pair plasma. Moreover, the idea of two flow structures was supposed. The jet is composed by a mildly relativistic sheath composed of $e^-/p^+$ and driven by magnetohydrodynamical (MHD) forces and an ultra-relativistic flow composed of $e^-,e^+$ pairs, which is responsible for most of the emission (see Fig.~\ref{Fig9}). We assume that the AGN is composed of a central rotating black hole (BH), with two different accretion disks: around the central BH for radius $R<10 r_g$ an advection-dominated accretion flow 
(ADAF) ($r_g$ is the Schwarzschild radius, $r_g=2MG/2c^2$, $M$ is the BH mass, $G$ the gravitational constant and $c$ the speed of light) and an external standard accretion disk (SAD). The SAD emits photons mainly in ultra-violet and X rays and the ADAF emission is between the radio and gamma-ray range~\cite{Oka_2003}. 

The mechanisms of $e^-,e^+$ pair plasma production and acceleration are not well understood, however we will assume that the pair plasma is produced from the Breit-Wheeler process. As it is shown in Fig.~\ref{Fig9}, the gamma-rays emitted from the ADAF collide along the BH rotation axis. Once the pair plasma is created by the BW process, the anisotropy of the high energy photons emitted from the accretion disk can transfer a strong momentum to the pair plasma via the inverse Compton process. This mechanism is know as Compton rocket~\cite{Odel_1981}. We propose to study direct pair plasma creation and acceleration by the BW process, and show that this flow can be produce with high Lorentz factor $\Gamma_b\sim 10$ without the requirement of the Compton rocket mechanism. 

The characteristics of the photon energy distribution emitted from the ADAF was calculated in Ref.~\cite{Oka_2003}, the photons are emitted from the radio to gamma-ray range, with an energy cuttoff at 4~MeV (see Fig.~2 in Ref.~\cite{Oka_2003}). From the pair threshold production definition, only photons between $0.065-4$~MeV energies can participate to the BW pair production (see Sect.~\ref{Sect1}). We propose to study the photon beams collision for different angles $\theta_p$, in two different cases: for 4~MeV-4~MeV and 4-0.5~MeV photon beam energies. For the second situation, 0.5~MeV has been chosen to highlight the beaming effect and the consequence of energy difference between the two photon beams.   
\begin{figure*}
\centerline{\includegraphics[width=16cm]{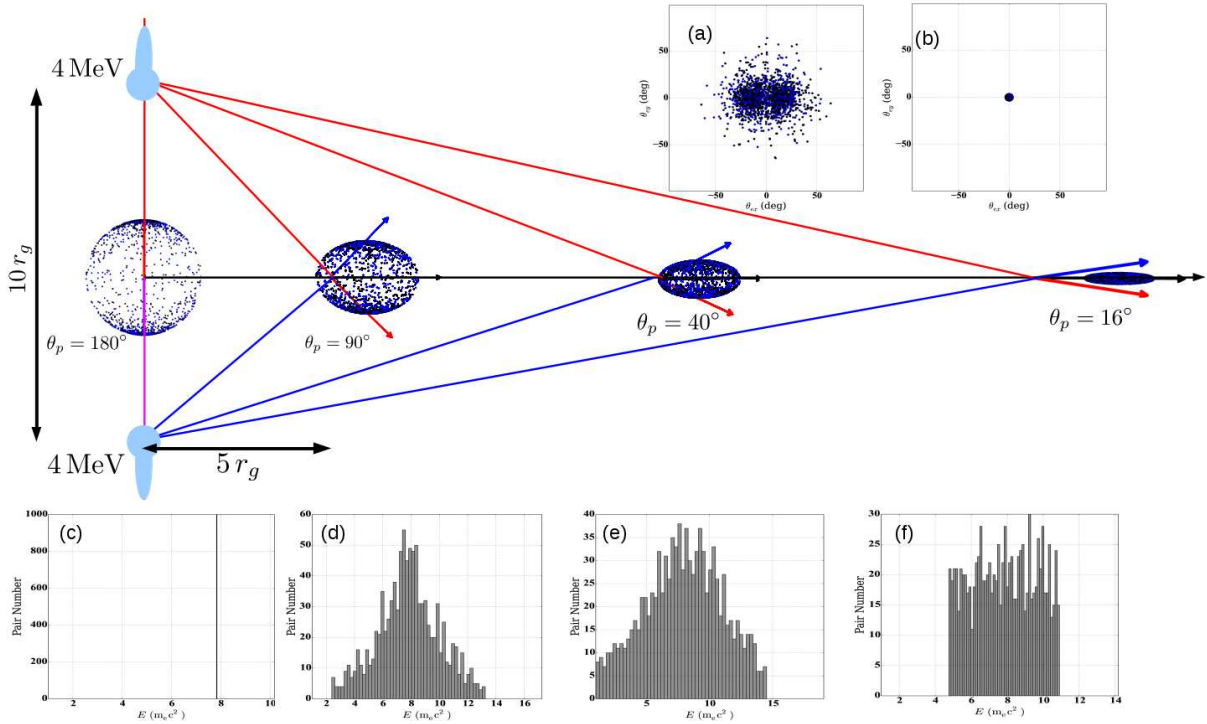}}
\caption{\label{Fig10} Schematic picture of the AGN, two beam sources of equal energy 4~MeV are emitted symmetrically by the ADAF and collide on the axis. For $\theta_p=180^{\circ}; 90^{\circ}, 40^{\circ}, 16^{\circ}$, the momenta for each photon beam collisions are plotted. (a)-(b) show angular pairs distribution when beaming condition is satisfied. (c)-(f) show the energy distribution for the different $\theta_p$ values.}
\end{figure*}

\begin{figure*}
\centerline{\includegraphics[width=16cm]{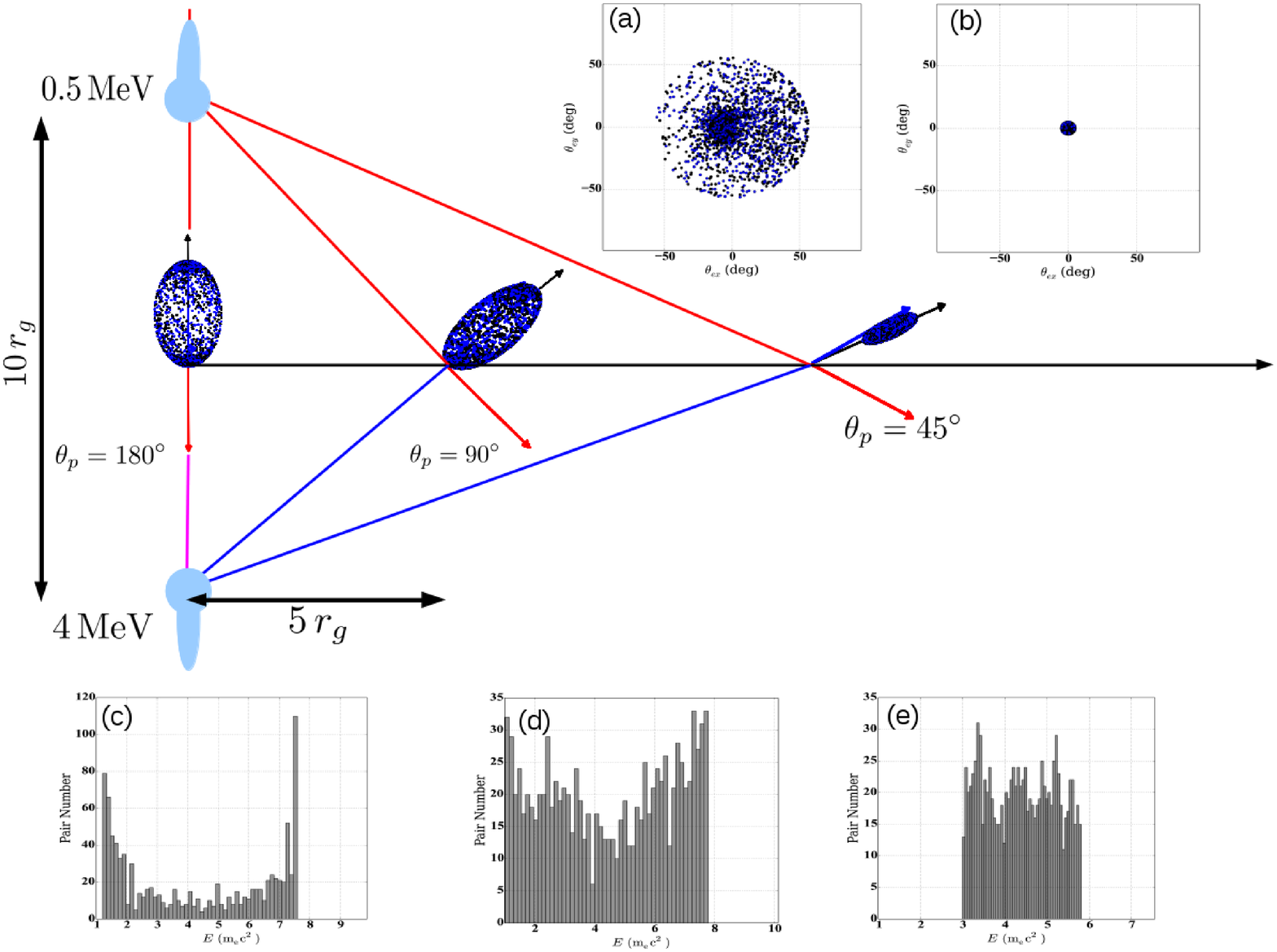}}
\caption{\label{Fig11}
Same situation than as in Figure~\ref{Fig10}, but with the two photon beam sources of energies 4~MeV-0.5~MeV colliding on the symmetry axis for $\theta_p=180^{\circ}; 90^{\circ}, 45^{\circ}$.}
\end{figure*}

Figure~\ref{Fig10} shows for $\theta_p=180^{\circ}, 90^{\circ}, 40^{\circ}, 16^{\circ}$ the pair momenta distributions for 4~MeV-4~MeV photon beams collision. From the pair beaming criteria, only the pairs corresponding to $\theta_p \in [14.7^{\circ},41.9^{\circ}]$ are beamed. For $\theta_p=180^{\circ}$ the pairs are emitted toward the photon beam directions and the energy distribution is mono-energetic (see Fig.~\ref{Fig10}c), all the pairs are emitted with Lorentz factor $\Gamma_b = 8$. For $\theta_p=90^{\circ}$ most of the pairs are emitted in CM frame velocity direction, however, a small fraction are emitted toward the BH. The pair energy distribution is peaked around $\Gamma_b\sim 8$, with a maximum $\Gamma_b\sim 13$ (see Fig.~\ref{Fig10}d). For $\theta_p=40^{\circ}$ the pairs are beamed, all the pairs propagate along the CM frame velocity direction. The pair angular distribution is shown in Fig.~\ref{Fig10}a and the inner beam pair 
structure is due to the BW differential cross section (see previous section). For the beaming angle, the pair beam achieves the maximum energy: $\Gamma_b\simeq 15$. Between $\theta_p=40^{\circ}-16^{\circ}$ the pair beaming increases (see Fig.~\ref{Fig10}b) and at the same time the pair energy distribution range decreases (see Fig.~\ref{Fig10}f).

Figure \ref{Fig11} shows the pair distribution characteristics for 4~MeV-0.5~MeV photon beams collision. We represent the momenta distribution for $\theta_p=180^{\circ},~90^{\circ},~45^{\circ}$. Because of the different photon energies (see the previous section), the pair beam is emitted mainly along the higher photon energy direction. For $\theta_p=180^{\circ}$ the pair momenta distribution are preferentially emitted in the direction of the CM frame velocity, that is the direction of the maximum photon energy beam. The corresponding energy distribution in Fig.~\ref{Fig11}c shows that the energy distribution is no more a Dirac distribution. As in the previous case, the distribution is enlarged between $\Gamma_b~\sim~1-7.5$ and the pair beam is emitted preferentially at low and high energies. The beaming condition is satisfied for $\theta_p \in [42.4^{\circ},98.6^{\circ}]$, then all the pairs are emitted in the CM frame velocity direction. However, the pair beam axis is no longer on the AGN symmetry axis, but 
with an angle of $45^{\circ}$ from the symmetry axis. 
On Fig.~\ref{Fig11}a the pair angular distribution is peaked due to BW differential cross section effect. The energy distribution is more uniform and achieves the maximum extent, as we observe in the previous case, with $\Gamma_b~1-8$ (see Fig.~\ref{Fig11}d).  Finally, for $\theta_p=45^{\circ}$ the pair momenta distribution is pinched along the CM frame velocity direction (see Fig.~\ref{Fig11}b). The pair beam is emitted with an average angle of $\sim 20^{\circ}$ from the AGN symmetry axis and pairs are beamed inside $\pm 5^{\circ}$. The corresponding energy distribution varies between $\Gamma_b~\sim~3-6$ (see Fig.~\ref{Fig11}e).

In summary, the two different photon collision situations allow to observe that for $4-4$~MeV photon beam energies, the pair beam is aligned along the AGN symmetry axis, with the maximum Lorentz factor $\Gamma_b\sim 15$ when the beaming criteria is reached. The pair beam is more collimated due to the BW differential cross section effect. While in the $4-0.5$ MeV case the pair beam is off axis with a lower Lorentz factor $\Gamma_b < 8$ and is less collimated.

\section{Conclusions}\label{Sect6}
The pair beaming condition in BW process has been previously studied in Ref.~\cite{Ribeyre_2017}. In this paper, we go further to investigate in details the effect of the BW differential cross section on the pair beaming. We show that, for equal photon beam energies, this effect is weak, i.e. for $\sqrt s<3$, the pairs are emitted mainly on the bisector of the initial photon beam directions. However, for different photon beam energies, the pairs are beamed in the CM frame velocity direction. For $\sqrt s >3$, the effect the BW differential cross section becomes important. Two peaks in the pair angular distribution appear in the photon incident plane along the photon beam directions. In the case of equal photon energy beams, the two peaks are symmetric and for unbalanced energies more pairs are emitted toward the highest photon energy direction. The energy distributions are modified in consequence.

An application of the BW process to an astrophysical situation, in particular to AGN, shows that the pair beam achieved the maximum Lorentz factor when the beaming conditions are satisfied. For the 4~MeV-4~MeV photon collisions situation, Lorentz factors $\Gamma_b$ greater than 10 can be reached. In this case $\Gamma_b$ values obtained are close to the AGN jets observations and the BW differential cross section effect is important. For the 4~MeV-0.5 MeV photon collision situation, the pair beam is off-axis and the pair beam energy distribution is shifted to lower energies. Then, this study suggests that the BW process could be taken into account in initial conditions or in addition of the Compton rocket pair acceleration process. 

\ack
We acknowledge the financial support from the French National Research Agency (ANR-17-CE30-0033-01) - TULIMA Project. This work is partly supported by the Aquitaine Regional Council. 


\end{document}